\documentclass[11pt,a4paper]{article}
\usepackage[british]{babel}
\usepackage{amsmath,amssymb,natbib,multicol}

\newcommand{\mkG}{\,\mu{\rm G}}
\newcommand{\nG}{\,{\rm nG}}

\newcommand{\tmop}[1]{\ensuremath{\operatorname{#1}}}

\def\jnl@style#1{{\rmfamily#1}}%
\def\jref@jnl#1{{\jnl@style#1}}%

\newcommand\na{\jref@jnl{NewA}}%
\newcommand\aj{\jref@jnl{AJ}}%
\newcommand\araa{\jref@jnl{ARA\&A}}%
\newcommand\apj{\jref@jnl{ApJ}}%
\newcommand\apjl{\jref@jnl{ApJ}}%
\newcommand\apjs{\jref@jnl{ApJS}}%
\newcommand\ao{\jref@jnl{Appl.~Opt.}}%
\newcommand\apss{\jref@jnl{Ap\&SS}}%
\newcommand\aap{\jref@jnl{A\&A}}%
\newcommand\aapr{\jref@jnl{A\&A~Rev.}}%
\newcommand\aaps{\jref@jnl{A\&AS}}%
\newcommand\azh{\jref@jnl{AZh}}%
\newcommand\baas{\jref@jnl{BAAS}}%
\newcommand\jrasc{\jref@jnl{JRASC}}%
\newcommand\memras{\jref@jnl{MmRAS}}%
\newcommand\mnras{\jref@jnl{MNRAS}}%
\newcommand\nar{\jref@jnl{NewAR}}%
\newcommand\pra{\jref@jnl{Phys.~Rev.~A}}%
\newcommand\prb{\jref@jnl{Phys.~Rev.~B}}%
\newcommand\prc{\jref@jnl{Phys.~Rev.~C}}%
\newcommand\prd{\jref@jnl{Phys.~Rev.~D}}%
\newcommand\pre{\jref@jnl{Phys.~Rev.~E}}%
\newcommand\prl{\jref@jnl{Phys.~Rev.~Lett.}}%
\newcommand\pasp{\jref@jnl{PASP}}%
\newcommand\pasj{\jref@jnl{PASJ}}%
\newcommand\qjras{\jref@jnl{QJRAS}}%
\newcommand\skytel{\jref@jnl{S\&T}}%
\newcommand\solphys{\jref@jnl{Sol.~Phys.}}%
\newcommand\sovast{\jref@jnl{Soviet~Ast.}}%
\newcommand\ssr{\jref@jnl{Space~Sci.~Rev.}}%
\newcommand\zap{\jref@jnl{ZAp}}%
\newcommand\nat{\jref@jnl{Nature}}%
\newcommand\iaucirc{\jref@jnl{IAU~Circ.}}%
\newcommand\aplett{\jref@jnl{Astrophys.~Lett.}}%
\newcommand\apspr{\jref@jnl{Astrophys.~Space~Phys.~Res.}}%
\newcommand\bain{\jref@jnl{Bull.~Astron.~Inst.~Netherlands}}%
\newcommand\fcp{\jref@jnl{Fund.~Cosmic~Phys.}}%
\newcommand\gca{\jref@jnl{Geochim.~Cosmochim.~Acta}}%
\newcommand\grl{\jref@jnl{Geophys.~Res.~Lett.}}%
\newcommand\jcp{\jref@jnl{J.~Chem.~Phys.}}%
\newcommand\jgr{\jref@jnl{J.~Geophys.~Res.}}%
\newcommand\jqsrt{\jref@jnl{J.~Quant.~Spec.~Radiat.~Transf.}}%
\newcommand\memsai{\jref@jnl{Mem.~Soc.~Astron.~Italiana}}%
\newcommand\nphysa{\jref@jnl{Nucl.~Phys.~A}}%
\newcommand\physrep{\jref@jnl{Phys.~Rep.}}%
\newcommand\physscr{\jref@jnl{Phys.~Scr}}%
\newcommand\planss{\jref@jnl{Planet.~Space~Sci.}}%
\newcommand\procspie{\jref@jnl{Proc.~SPIE}}%

\begin{document}

\title{Probing magnetic fields with Square Kilometre array and its precursors}

\author{Subhashis Roy\footnote{National Centre for Radio Astrophysics, TIFR, Ganeshkhind, Pune 411007}, 
Sharanya Sur\footnote{Indian Institute of Astrophysics,II Block, Koramangala, Bangalore 560034}, 
Kandaswamy Subramanian\footnote{Inter University Centre for Astronomy \& Astrophysics, Ganeshkhind, Pune 411007}, 
Arun Mangalam\footnote{Indian Institute of Astrophysics,II Block, Koramangala, Bangalore 560034},\\
TR Seshadri\footnote{Dept of Phys. and Astrophys., University of Delhi, New Delhi, Delhi 110007} \& Hum
Chand\footnote{Aryabhatta Research Institute of Observational Sciences, Manora Peak, Nainital 263002}}

\maketitle

\begin{abstract}
Origin of magnetic fields, its structure and effects on dynamical processes in stars to galaxies are 
not well understood. Lack of a direct probe has remained a problem for its study. The first phase of 
Square Kilometre Array (SKA-I), will have almost an order of magnitude higher sensitivity than the 
best existing radio telescope at GHz frequencies. 
In this contribution, we discuss specific science 
cases that are of interest to the Indian community concerned with astrophysical turbulence and 
magnetic fields. The SKA-I will allow observations of a large number of 
background sources with detectable polarization and measure their Faraday
depths (FDs) through the Milky Way, other galaxies and their circum-galactic
mediums. This will probe line-of-sight 
magnetic fields in these objects well and provide field configurations. Detailed comparison of observational 
data (e.g., pitch angles in spirals) with models which consider various processes giving rise to field 
amplification and maintenance (e.g., various types of dynamo models) will then be possible. Such 
observations will also provide the coherence scale of the fields and its random component through 
RM structure function. Measuring the random component is important to characterise turbulence in the 
medium. Observations of FDs with redshift will provide important information on magnetic field evolution 
as a function of redshift. The background sources could also be used to probe magnetic fields and its 
coherent scale in galaxy clusters and in bridges formed between interacting galaxies. Other than FDs, 
sensitive observations of synchrotron emission from galaxies will provide complimentary information 
on their magnetic field strengths in the sky plane. The core shift measurements of AGNs can provide 
more precise measurements of magnetic field in the sub parsec region near the black hole and its
evolution. The low band of SKA-I will also be useful to study 
circularly polarized emission from Sun and comparing various models of field configurations with 
observations.
\end{abstract}

\section{Introduction}

Magnetic fields are detected in almost every astrophysical object in the Universe. 
Starting from stars like our Sun, the interstellar medium of nearby spiral galaxies 
and the intracluster medium of galaxy clusters, all host dynamically important 
magnetic fields. As a consequence, magnetic fields play a wide variety of roles 
ranging from controlling present day star formation where it possibly determines the 
typical masses of stars and the initial mass function, regulating the propagation 
and confinement of cosmic rays to the launching and maintenance of galaxy outflows. 
In disk galaxies, magnetic fields contribute significantly to the total pressure of the 
interstellar gas while outflows from galaxies control the magnitude and structure of 
magnetic fields in galactic halos 
and could be responsible for magnetising
the intergalactic medium.

Magnetic fields can be probed chiefly by four techniques : (i) synchrotron emission in the radio 
band (both in the continuum and polarized components), which provides its field strength and 
orientation in the sky plane. (ii) Zeeman effect, which independently measures its strength in 
cold gas clouds, and (iii) Faraday rotation, which provides the line of sight averaged fields. 
Polarization angle provides its orientation in the sky plane. Other than the above, non-relativistic 
electrons in presence of magnetic fields produces (iv) cyclotron emission. Given the typical magnetic 
field strengths in stellar bodies, such narrow band circularly polarized emission could be detected at 
radio frequencies. This technique can directly yield the magnetic field strength, and is often used to 
study magnetic fields in Sun and other stars.
On the theoretical front, conservation of 
magnetic helicity has now been recognised as a crucial ingredient in amplifying magnetic 
fields to saturated strengths \citep{BS05}. Nonlinear models of magnetic field evolution based 
on this conservation law is already in place for probing galactic magnetic fields 
\citep[e.g.,][]{SSSB06} and efforts are on to incorporate it in the solar context 
\citep[see][for a review]{Charbonneau14}.

In spite of these important developments, our understanding of how magnetic fields are 
amplified and maintained in various astrophysical objects is far from clear. The 'Origin of 
Cosmic Magnetism' is one of the main themes of research of the upcoming
Square Kilometer Array which will have a collecting area of more than an order of magnitude 
higher than the existing largest radio telescope (GMRT). In the first stage, referred to as 
SKA-I, it is envisaged to have a sensitivity of almost an order of magnitude higher 
\footnote{\small {https://www.skatelescope.org/wp-content/uploads/2014/03/SKA-TEL-SKO-0000007\_SKA1\_Level\_0\_Science\_RequirementsRev02-part-1-signed.pdf}} than 
the Jansky Very Large Array (JVLA) and is expected to finish by early next decade. 

Given this rapid advancement in observational progress driven by the construction of the 
SKA, we discuss in this article, some of the issues associated with cosmic magnetic fields 
which are of interest to the Indian community that we aim to address with the SKA-I.

\section{Magnetic fields in galaxies}

\subsection{Challenges in measuring magnetic fields in galaxies}

Synchrotron emission and its polarization are useful tracers of magnetic fields, but are dominantly detected 
in regions where the density of cosmic rays and associated relativistic electrons are relatively high (i.e., near 
star forming regions), or where the magnetic field is stronger. However, certain regions of interest for magnetic 
field studies are far from star formation regions and supernova remnants (e.g., inter-arm regions in galaxies), 
and these regions could have to be excluded for magnetic field studies, which could bias the results. Moreover, 
this method yields magnetic fields in the sky plane under the assumption 
of energy equipartition between cosmic rays and magnetic fields (see, e.g., \citet{BASU2013}). 
However, this method suffers from assumptions of equipartition between different components of ISM and they 
can deviate from equipartition values in certain conditions and below certain size scales.

A much more pervasive probe of interstellar magnetic fields is by Faraday rotation. Due to this effect, 
magneto-ionic ISM causes the position angle of a linearly polarized wave to rotate. For an electromagnetic 
wave emitted 
from a source at position $r$, with intrinsic
position angle $\theta_0$ at a wavelength $\lambda$, the detected position angle is
\[ \theta' = \theta_0 + \tmop{FD}\,\lambda^2 \]
In this expression, the Faraday depth (FD), in units of rad m$^{- 2}$, is defined by
\[ \tmop{FD} = K \int B \cos (\phi) n_e d l \] where $K$ =0.81 rad m$^{- 2}$ pc$^{- 1}$ cm$^{- 3}$ $\mu G^{- 1}$, $B$, 
$\phi$ and $n_e$ are the magnetic field strength (in $\mu$G), inclination of the magnetic field vector to the line of sight 
and number density of thermal electrons (in cm$^{-3}$), respectively. The integral is carried out along the line of sight 
from source at $r$ to the the observer.
When the emission originates from a single FD, it is equivalent to
the Faraday rotation measure (RM).
This typically happens if there is a single foreground magneto-ionic medium inducing the
Faraday rotation of a background source. However, in general polarized synchrotron radiation
arises from the same volume that is also inducing Faraday rotation, in which case the net
polarized intensity is a superposition of the emission from various FD's with corresponding
Faraday rotation. Then the observed RM (got by fitting a $\lambda^2$ law to the position
angle of the polarization) will not be equivalent to a single FD.

Multi-wavelength observations of background polarized sources can directly yield the FD along the line of 
sight. However, even in simple cases with a single FD, it does not yield a
value for $B$. The sign of the FD can provide information on the direction of
the magnetic field. When, electron density along the line of sight is known,
the mean amplitude 
of the magnetic field strength along the line of sight can be directly inferred. In external galaxies, electron 
density is frequently estimated from H$_{\alpha}$ emission. However, this is easily absorbed by dust, and 
extinction correction due to dusts at far-IR wavelengths is problematic due to lower resolution and sensitivity 
of existing observations in this band. 
Moreover we note here that ${\rm H}_{\alpha}$ emission is dominated by clumpy ionized gas (HII regions), 
while the FD and the RM is dominated by the diffuse ionized gas. Thus, ${\rm H}_{\alpha}$ 
intensity can only be used to provide a model of the electron density distribution in galaxies.
Nevertheless, in the modern era, especially with the launch of Herschel, the resolution at far-IR wavelengths is 
approaching the resolution of radio telescopes. Reliable measurements of electron densities in external galaxies to 
convert FDs to a line of sight averaged value of $B$ should then be possible in SKA-I era. 

In our galaxy, pulsar dispersion measure can provide the line of sight electron density, which can be used to find 
an average value of B$_{||}$ from its FD
(here B$_{||}$ is the line of sight magnetic field component). 
However, even in this case, there could be complications due to correlations 
between B$_{||}$ and n$_e$. Since these are not easily accounted for, getting accurate measurements of B$_{||}$ by this 
method is difficult even at higher radio frequencies where depolarization is insignificant.

\subsection{Survey of background sources for polarized emission with SKA (RM grid)}

As discussed above, the drawbacks of measuring magnetic fields through indirect means cannot go 
away with new telescope. However, SKA-I with 2 orders of magnitude of higher sensitivity than the old 
VLA will sample the magnetic fields in galaxies through observations of background sources much 
better than the existing observations.  For example, with the (old) VLA, one could find measurable 
Faraday rotation towards only $\sim$1 source per square degree with an integration time of $\sim$10 
minutes per source. This seriously limits the number of measurements available in a given region and 
affects the reliability of results derived in complex regions in the Galactic plane (see, e.g.,\citet{ROY2008}). 
Typical angular size of nearby ($\sim$10 Mpc) spiral galaxies is $\lesssim 10'$. Therefore, even with integration 
time of an hour per background sources, there would only be a few background sources with measurable 
polarization/RM seen through nearby galaxies. Therefore, deciphering magnetic field structures in these 
galaxies is very challenging by this method with the existing radio telescopes.

As discussed in \citet{BECK2004}, an appropriate observing wavelength for detecting large numbers of 
polarised sources is $\sim$GHz. This provides a large field of view, without introducing severe internal depolarization 
effects which will prevent FDs from being measured in many extragalactic background sources.  With an 
angular resolution of better than 1$''$, if typical extragalactic sources are polarized at $\sim$3\%, this will
allow measurable FDs towards $\sim$1000 sources per square degree in the sky with an integration time 
of $\sim$1 hour per field (assuming the field of view to be about a degree). This will allow disentangling of 
local FD effects from the systematic ones in our Galaxy, and allow a detailed map of magnetic fields in the Galaxy.  
If one observes nearby galaxies (size $\sim$10$'$) with SKA-I at 1.4 GHz for $\sim 1$ hour, several tens of
background sources with measurable FDs will be seen through them. This will reduce existing sampling bias to 
a large extent.
To improve sensitivity of FD observations, steep spectrum of background sources need to be considered
while minimising significant depolarization. SKA1-MID band 4 from 2.8-5 GHz is well suited. It provides detection 
of minimum FD of a few tens of rad m$^{-2}$ and can detect FD upto about 500 rad m$^{-2}$ \citep{BECK2015B}. 
Typical sensitivity would be about 0.2 $\mu$Jy/beam with integration time of 12 hour.
As a caveat, we note here that most of the polarised AGN's required to measure the foreground RM have 
complicated internal structure \citep{AGF16, Kim+16}. Measuring the internal Faraday structure of these sources 
could require additional observations both in the lower frequency band around 1 GHz and higher frequency band 
($\sim 8$ GHz) to identify complicated internal structures in FD space. These sources may either be avoided or 
suitably modeled before using them as probes of the intervening medium.

\subsection{Rotation measure synthesis and magnetic field tomography}

Multifrequency observations of polarized sources in the past have been used exclusively to determine their 
RMs. A linear fitting of measured polarization angles to $\lambda^2$ was used for the above. However, the 
above relation is only true when there is a single compact source in the synthesised beam of a radio telescope 
and there is no line of sight superposition with any other Faraday screen. When there are more than one source 
in a telescope beam or the source is extended, the simple linear relationship of FD with $\lambda^2$ breaks down. 
This is due to depolarization and was first discussed in detail by \citet{BURN1966}. He also introduced a Fourier 
transform relationship between complex polarized surface brightness (P) and FD ($\phi$). \citet{BRENTJENS2005}
have further considered the case of limited sampling of $\lambda^2$ space and on constant emission spectrum. 
Many of the modern radio astronomy telescopes have large bandwidth and a large number of spectral channels 
across the observing band. For such cases, Fourier inversion of the measured complex polarized emission as
a function of frequency channels is possible. RM synthesis by \citet{BRENTJENS2005} can be employed to recover 
the RMs of individual Faraday screens when several of them are located along the same line of sight. The
$\lambda^2$ coverage of any given observations is incomplete, and a sampling function called the RM spread 
function (RMSF) is defined in the Faraday domain. 
It is also necessary to make an assumption about P for negative values of $\lambda^2$.
The Faraday spectrum can be deconvolved using RMSF
through {\it RMCLEAN} \citep{HEALD2009}. In certain situations, the technique may not work well and can be 
complemented by fitting in the Q, U domain \citep{O'SULLIVAN2012}. Longer radio wavelengths are required for 
a high FD resolution $\delta\phi=2.\sqrt{3}/\Delta\lambda^2$ (where, $\Delta\lambda$ is the wavelength 
coverage). The maximum FD that can be measured is given by $|\phi_{max}|=\sqrt{3}/\delta\lambda^2$ 
(where, $\delta\lambda$ is the channel width) \citep{BRENTJENS2005}. Typical FD in the tomography 
studies varies from a few to a few hundred rad m$^{-2}$. This requires broad frequency coverage with the lower
part of the observing band going down below 1 GHz. Observations with SKA survey band 2 from 650 MHz to 
1.7 GHz would be very useful. The lowest frequency of SKA1-MID is proposed to be 950 MHz and will also be 
useful for the purpose. 
The upgraded Giant Metrewave Radio Telescope (GMRT) will also have an observing band from about 
600$-$900 MHz along with the existing 1$-$1.4 GHz band. It should be possible to use these bands of GMRT 
to gain experience with RM synthesis before SKA1 survey band 2 is operational.  We note that Ionospheric
Faraday rotation causes significant problems in calibrating radio polarization data at metre wavelengths. 
Therefore, polarization observations at frequencies below 500 MHz could be difficult to calibrate with the GMRT 
for most sources (except pulsars) with weak polarized emission.

At lower frequencies, internal depolarization is present everywhere in galaxies and causes a frequency dependent 
modulation of the degree of polarization. Using the methodology as described above, it provides a great deal of
information about the physical conditions in ISM of galaxies. Through detailed modeling of depolarization, 
3-dimensional structure of the Faraday screen and magnetic fields in the ISM of galaxies can be unveiled in great 
details \citep{HEALD2015}. This is complementary to higher frequency RM grid observations discussed earlier, 
which suffers little from depolarization.

\subsection{Magnetic fields in the sky plane}

Magnetic fields in the sky plane is not probed by the Faraday RMs described above. 
Synchrotron emission intensity under the assumption of energy equipartition of cosmic rays
and the magnetic fields 
provides an estimate of the magnetic fields in the sky plane. This will provide complimentary information of what 
Faraday RM provides along our line of sight and is important for probing magnetic fields in galaxies which are 
almost face-on. In these objects, the systematic fields are likely oriented perpendicular to our line of sight (along 
the galactic plane). Probing magnetic fields by FD alone will fail to give complete information about these systematic 
fields. In radio, these fields are probed by synchrotron emission. Despite the limitation due to assumption of energy 
equipartition it does provide a value for the total magnetic fields in the sky plane, when the thermal emission is 
removed from the total emission and the synchrotron spectral index is measured. Moreover, the large scale structure 
of the field, and its symmetries (axi-symmetric, bi-symmetric etc.) can be constrained by observing the synchrotron 
polarisation angle. 
However, it must be noted that apart from the 'ordered' field, polarization can also be produced by an 'anisotropic' field 
that frequently reverses its direction incoherently. Both these types of fields may produce the same degree of polarization. 
A map of Faraday rotation measure (which will show large-scale coherence only for the ordered field) is therefore essential 
to distinguish between the two.
SKA-I MID will have large bandwidths which will allow to simultaneously fit for thermal emission and spectral index of 
non-thermal fraction (especially between 1-1.8 GHz) across whole field of view with significant emission. Many of the spiral 
galaxies at distances of $\sim 10$ Mpc are well sampled by the present day radio telescopes like VLA and the GMRT. 
SKA-I with a few times better resolution and more than an order of magnitude higher sensitivity than the JVLA can easily 
extend the distance scale to a few times larger distance to image these galaxies. This will allow to increase the sample size 
by almost two orders of magnitude.

\section{Scientific applications of the magnetic field measurements}

\subsection{Origin of magnetic fields in galaxies}

There are two competing paradigms to explain the observed magnetic fields in spiral galaxies. One of them 
asserts that the observed structures evolved from a {\it primordial magnetic field} acted upon by differential rotation. 
The other explanation is that these magnetic fields resulted from some form of {\it dynamo} action by 
tapping the energy in turbulent fluid motions and rotational shear within the galaxy to amplify magnetic fields from 
{\it seed fields}, until the Lorentz forces back-react on the flow to saturate the field growth. 
Depending on the scale over which the resulting magnetic field is generated, they are classified as 
'Large-scale (Mean-Field)' and 'Small-scale (Fluctuation)' dynamos. Large-scale dynamos generate magnetic fields 
that are spatially coherent on scales much larger than the energy carrying scale of turbulence, while fields generated 
by small-scale dynamos are coherent on scales on or below the energy carrying scale of turbulent motions.
In stars and galaxies, these may often be both operative and then an important question is how they
interact and come to terms with each other \citep{BHAT2016}.

There are however two 
main issues that work against the idea of a primordial origin - a) in the absence of turbulent diffusion, twisting of a 
primordial field by differential rotation would produce a tightly wound spiral with the field alternating on very small radial 
scales $\sim 0.1\,{\rm kpc}$ and producing a magnetic pitch angle $p =-1^{\circ}$ \citep{Shukurov04}, that is impossible 
to reconcile with the observed pitch angles of $-15^{\circ}$ (see later subsections below). The tight winding could 
be alleviated by turbulent diffusion expected to occur naturally in turbulent environments but then b) this would lead 
to the dissipation of the large-scale field on a timescale $\sim 5\times10^{8}$ yr, which is a fraction of the galactic lifetime 
\citep{Shukurov04, Brand15}. In light of these difficulties it therefore seems inevitable that irrespective of the origin of 
the 'seed' field, some {\it in situ} mechanism is necessary to amplify the fields to observed strengths and sustain them 
over the galactic lifetime.

In this context, the {\it dynamo} mechanism appears to be the most promising candidate to explain the 
observed magnetic fields \citep{Brandenburg2005}. 
Even though, the 
mean-field
galactic dynamo theory has been able to predict some 
of the essential features of magnetic fields in nearby spirals (e.g., the observed quadrupolar symmetry of the regular 
magnetic field w.r.t the Galactic equator in the Milky Way \citep{Frick+01}) , a complete nonlinear model of the 
galactic dynamo taking into account the intricacies of star formation and gas dynamics is still in its infancy. We 
postpone a discussion of some of these issues and how
we plan to address them with the SKA to the later subsections.

\subsection{The Milky Way}

Milky Way, due to its proximity, is an ideal test bed for studying magnetic fields in spiral galaxies.  We do have 
detailed studies of the various constituents of the gaseous ISM in the Galaxy which could help in determining
the magnetic field structure and strength. Magnetic fields in discrete objects like molecular clouds, supernova 
remnants, HII regions, planetary nebulae and in small scales ($<$10 pc) are best explored in detail in the Galaxy. 
Large scale reversals in arms of the Galaxy can also be studied.  Our location in the Galaxy does however create 
difficulties in certain studies involving large angular scales (e.g., the halo).

Galactic magnetic fields are believed to be amplified and maintained by dynamos. 
The necessary energy could come from differential rotation of the Galaxy, turbulence in the 
ISM and cosmic rays. Regular fields are thought to be amplified by the mean-field alpha-omega 
dynamo. As discussed further in Sect.~3.5, the dynamo mode most easily excited is axisymmetric 
and has even parity with respect to the mid plane. The toroidal and radial component forms an
axisymmetric spiral pattern. In spherical objects, such as in galactic halos, it is expected to generate 
axisymmetric mode with odd vertical parity, in which the magnetic fields perpendicular to the plane pass 
the mid plane continuously, while the horizontal component reverses direction across the mid plane.
In the disk and halo of our Galaxy, this would result in mixed-parity modes \citep{SOKOLOFF1990}.  
In presence of a Galactic wind, the mixed parity mode can be sustained \citep{MOSS2010}.  
Apart from these simple magnetic field configurations, observations from existing sample of galaxies 
also show that realistic fields can have much more complicated configurations such as revealed in galaxies
with a bar or in galaxies
like NGC891 and NGC4565, which show an X-shaped pattern with vertical field components that increase with 
increasing height above and below the galactic plane and also with increasing radius. Compression and 
shearing by gas flows may further shape the fields and overrun the simplistic dynamo patterns predicted 
from theory.

As described in detail in \citet{HAVERKORN2015}, comparing the predictions of dynamo theory with the
current observational data of the Galactic magnetic fields is still difficult. There is still uncertainty on whether 
the large scale field in the Galaxy is axisymmetric \citep{VAN-ECK2011} or bisymmetric \citep{NOTA2010}. 
This could explain the apparent contradiction of butterfly pattern of rotation measures (RMs) in the inner 
Galaxy to indicating reversals of azimuthal field across the Galactic plane \citep{SIMARD-NORMANDIN1980} 
to lack of reversals of the field component when done locally \citep{Frick+01}. Some models of the existing 
observations seem to confirm the mixed parity dynamo modes \citep{JANSSON2012}.  Characterising the 
global structure is, however, quite difficult due to contamination by local structures like the local magnetised 
bubble \citep{SUN2010,HAVERKORN2015}.

The observations which lead to the above results are using existing RM grid observations of extragalactic sources, 
pulsars and diffuse synchrotron emission in the Galaxy. The models are hampered by low density of polarized sources, 
uncertain distance estimates of pulsars, and contamination by various discrete objects in the Galaxy. Source density 
is only good in the Solar neighbourhood, but cannot determine the global field structure due to insufficient data points \citep{STEPANOV2002} (see also \citet{HAVERKORN2015}).

As discussed earlier, RM grid observations with SKA1-MID will yield magnetic fields every few arc-min along $l$ and 
$b$.  This will enable detailed study of magnetic fields in our Galaxy. Pulsar RM surveys with dependable distance
estimate will also probe magnetic fields along many different directions. It is estimated that SKA1-LOW and SKA1-MID 
will discover $\sim$20,000 pulsars in the Galaxy \citep{SMITS2011,KRAMER2015}.  This will help to model the magnetic
fields in the Galaxy.

\subsubsection{Turbulence in the Galaxy: }

\citet{CROVISIOR1981,Green1993} measured the power spectra of the atomic
hydrogen (HI) intensity fluctuation in Milky Way, which was found to follow a power law with a power law 
index of 2.6 at length scales ranging a few pc to 100s of pc.  These scale invariant structures have been 
understood to be the result of supernova driven turbulence in the ISM \citep{ES04}. \citet{DUTTA2013} 
estimated the power spectrum of the HI intensity fluctuation from 18 external spiral galaxies. They found 
that the power spectra follow power laws at length scales ranging a few 100 pc to 10s of kpc with most
of the galaxies having a power law index of 1.6. This dichotomy in the power law index (2.6 for Milky Way 
compared to 1.6 for external spirals) is understood by \citet{DUTTA2009} as an effect of geometry, where 
at scales shorter than the scale height of the disk three dimensional structures are probed, compared to the 
two dimensional structures at larger scales. Owing to the lower sensitivity at high resolution their measurements 
were limited to  $> 400$ pc scales. These observation raise a few  questions: what drives the structures at 
scales of tens of kpc and if the same mechanisms also influence the small scales structures seen in our galaxy, 
are these structure a result of local instabilities like the spiral arms etc. 

Major inputs required to answer these questions from the observations are as follows. Investigating the link 
between the large scale column density fluctuations seen in external galaxies and the small scale fluctuations 
seen in our Galaxy needs a measurement of the power spectrum over a large range of scales. Firstly, present 
day observations are not deep enough such that we can probe the HI column density power spectrum over 
more than a decade of length scales. Moreover the baseline coverage drops significantly at larger baselines, 
resulting in low sensitivity at those baselines where the power spectrum signal itself is low. It has been seen that 
the nature of the HI velocity dispersion is significantly different inside the stellar disk of the galaxies compared to 
the larger galacto-centric distances \citep{TAMBURO2009}. The observations also probe line of sight velocity of HI. 
Estimating the HI velocity fluctuation power spectrum would give us the direct measure of the energies involved in 
the process and hence would help us understand the generating mechanisms. A couple of visibility based estimators 
of the HI line of sight velocity fluctuation power spectrum are proposed \citep{DUTTA2016}, however the present 
observational data lacks the required sensitivity for these measurements. SKA-I mid will have better baseline coverage 
and sensitivity for these power spectrum measurements. It will be possible to probe the power spectrum of HI column 
density selectively at different parts of the Galaxy using estimators like Tapered Gridded Estimator \citep{Samir2016}.

Even though the power spectrum of the RM (RM structure function) is known, it's interpretation
in terms of the magnetic field spectrum,
may not be entirely straightforward. As discussed earlier, the estimate of the FD depends on the integral of the line 
of sight magnetic field component and the electron density. If the medium is homogeneous, the FD can be simply 
interpreted as an integral of the magnetic field along the sight from the source to the observer. However, 
densities in a realistic interstellar medium are rarely homogeneous. In that case, the estimate of the FD will 
depend crucially on the correlation (or anti-correlation) between $n_{\rm e}$ and ${\rm B}_{\rm ||}$. 
There are also additional potential complications in determining the magnetic field spectrum.

Observed spectral indices of RM are 
much flatter than the Kolmogorov
type \citep{HAVERKORN2008}. Magnetohydrodynamical turbulence is also characterised by scales of energy injection 
and dissipation, mach number of its flow speed, and the pressure ratio of the gas to the magnetic field in plasma. 
Measured maximum scales of fluctuation was earlier claimed to be $\sim$100 pc \citep{LAZARYAN1990}. But, more 
recent observations suggest a scale an order of magnitude smaller \citep{HAVERKORN2008, IACOBELLI2013}. 
The other parameters of the MHD turbulence remains uncertain \citep{HAVERKORN2015}. The turbulence could be 
significantly driven by supernova explosions in the Galaxy, and it would imply the turbulence to be intermittent at least 
in small and intermediate scales in the ISM (see also \citet{ES04}). Sect. 3.5 provides more details on turbulence.

\subsubsection{Magnetic fields near in the central region and in discrete objects}

{\bf Galactic centre (GC)}: Studying magnetic fields in the closest central region ($\sim$100 parsecs) 
of our Galaxy with a high resolution is important to understand the same in the central regions of similar spiral 
galaxies. The GC field might be the result of local effects (including any local dynamo) and accretion of matter 
from the rest of the Galaxy. A wind from the central region might also be required to stabilise the Galactic magnetic 
field configuration, and \citet{CROCKER2011} found evidence for the existence of such a wind. Large uncertainties 
exist in determining the field strength in this region. From synchrotron emission, the estimated minimum energy 
magnetic field is $\sim$10 $\mkG$.  However, synchrotron spectral break indicates the strength is at least
$\sim$50 $\mkG$ \citep{CROCKER2013}. It is now established using the RM towards the GC magnetar that within 
about 0.1 pc from Sgr-A\*, magnetic field strength could be several mG \citep{EATOUGH2013}. Magnetic field 
strengths in dense molecular clouds vary from a fraction of a mG to several mG \citep{CHUSS2003}. The structure of 
the magnetic fields are also unclear. Using RM grid measurements towards background sources, \citet{ROY2005,ROY2008}
suggested the field pattern to be consistent with either bi-symmetric spiral structure or oriented along the central bar. 
They estimated a field strength of $\sim$20 $\mkG$. However, \citet{NOVAK2003,LAW2011} interpret the polarization 
in the radio synchrotron filaments as indication of a poloidal (vertical to Galactic plane) field in the diffuse ISM.  
The discovery of Fermi bubbles have also raised questions on their origin. They could form due to occasional formation 
of jets due to the central black hole, or from stellar activity in this region \citep{CROCKER2011,CARRETTI2013}. Finding 
the Magnetic field structure will be important as it is likely to be linked to the maintenance of dynamo and formation of 
bubble like structure through stellar winds \citep{HAVERKORN2015}. As discussed earlier, RM grid observations with 
SKA1-MID will yield magnetic fields every few arc-min along $l$ and $b$.  This will enable detailed study of magnetic 
fields in the region. Observations above 5 GHz may be more suited to avoid significant depolarization due to high RMs 
in the region.

{\bf Molecular clouds:} Stars generally form from the collapse of molecular clouds. Magnetic fields could play an 
important role \citep{LI2014} in such a process. However, due to lack of accurate measurements its role has not been 
properly understood. Typically, magnetic fields in clouds are measured using Zeeman splitting of HI, OH, CN lines or 
masers, which provides the field strength along line-of-sight. Other methods including dust induced polarization are 
difficult due to inherent dependency on models. An alternative method is to use the intensity of in-situ synchrotron 
radiation from these clouds \citep{ORLANDO2013}. Since magnetic fields in molecular clouds could be much larger 
than the typical ISM \citep{CRUTCHER1999}, cosmic ray electrons while penetrating inside molecular clouds should 
generate synchrotron emission in presence of magnetic fields \citep{MARSCHER1978,DICKINSON2015}. Since cosmic 
ray spectrum in the Galaxy is known \citep{ACKERMANN2012}, and its density variation across the Galaxy has been 
studied, any detection will provide the total magnetic field which includes the turbulent component. Zeeman splitting is 
only sensitive to the regular magnetic fields. Polarized synchrotron emission could provide the ratio of regular versus 
the anisotropic random 
component \citep{DICKINSON2015}. There have been many observations to detect in-situ synchrotron emission 
from molecular clouds \citep{DICKINSON2015}, but they were not successful
\citep{YUSEF-ZADEH2013,YUSEF-ZADEH2007}. Clear detections and measuring magnetic fields 
in clouds allow us to infer the relative contribution of  gravitational, kinetic and magnetic energy densities in dense 
molecular clouds. It would enable testing star formation models involving ambipolar diffusion and turbulence \citep{CRUTCHER2012,LAZARIAN2012}.The effective brightness temperature of molecular clouds at 408 MHz using 
the cosmic ray flux model has been tabulated by \citet{DICKINSON2015} (see their Table-1), which shows that SKA1 
would easily detect many molecular clouds through their in-situ synchrotron emission and enable measuring total magnetic 
fields at different parts of the clouds. The dense cores of the clouds are typically less than 0.1 pc in size, subtending an 
angle $\sim 1 ^{''}$ at distances of the $\sim$kpc.  This matches well with SKA1-MID resolution and sensitivity. Estimated 
flux densities of $\sim$mJy can easily be detected at GHz frequencies.

{\bf Young stellar objects (YSOs):} YSOs quite often produce jets, but the launching mechanism of these jets and the 
role of magnetic fields in their production is ill understood \citep{RAY2009}.  One expects that active accretion to the
central object fuels these jets, where magnetic field could provide support for stabilising the magnetised jets 
\citep{HAVERKORN2015}. Synchrotron emission from most of these objects is quite weak, and SKA1 at frequencies 
of a GHz or so would be well suited to detect such faint small angular sized jet emission.

{\bf Supernova remnants (SNRs):} SNRs are the main drivers of turbulence and cosmic rays in the ISM. Magnetic 
fields in SNRs can be enhanced to mG strengths due to shock induced compression. It is believed that magnetic fields 
in SNRs carry the signature of large scale Galactic magnetic fields \citep{KOTHES2009}. There is also some evidence 
for toroidal fields believed to be caused by the stellar wind of the progenitor \citep{HARVEY-SMITH2010}. Measuring 
magnetic fields in a large sample of SNRs will shed light on any relationship of local Galactic magnetic fields and/or 
the progenitor fields in the evolution of the SNR and its magnetic fields \citep{HAVERKORN2015}. A knowledge of 
magnetic fields in SNRs will also help to compare the radio and high energy emissions from SNRs. As discussed
earlier, RM grid observations will yield magnetic fields every few arc-min along the shell of SNRs. This will enable 
detailed study of variation of magnetic fields around SNRs.

{\bf HII regions:} Magnetic fields in HII regions are also expected to be dependent on local Galactic magnetic fields. 
Magnetic fields could also affect the dynamics in HII regions. Studying a large no. of such objects over a large 
range of densities with RM grid could unveil any such dependence.

{\bf Faraday screens:} These are structures observed through polarization properties in the ISM, where no total intensity 
emissions could be identified \citep{HAVERKORN2003,Shukurov2003}. Better knowledge on their properties could 
arise from other studies involving polarization observations with SKA1 both through RM grid and Faraday tomography. 
\subsection{Nearby Galaxies:} 

\subsubsection{Coupling of magnetic fields with ISM}

The ratio of radio and far-infrared (FIR) flux density is known to be well correlated among galaxies. Radio emission 
in galaxies at GHz frequencies is mostly generated from non-thermal process, while the FIR emission is 
thermal in nature and is produced by dust. The relationship holds good over more than five orders of
magnitude \citep{CONDON1992}.  The relationship also does not evolve between local galaxies to galaxies
at redshift up to three \citep{MURPHY2009}. There are several models to explain the correlation. In the calorimeter 
model, young massive stars are considered as a source of IR (through reprocessing of stellar emission) and radio 
emission (through supernovae). However, \citet{NIKLAS1997} found a tight and non-linear global correlation between 
gas density and magnetic fields under equipartition assumption. This correlation also holds locally in galaxies, which 
is explained by close coupling of gas and magnetic fields with the form $B \propto \rho^k_{gas}$ 
\citep{HELOU1993,THOMPSON2006}. However, the slope of the correlation varies between arm and inter-arm regions. 
Depending on the observing frequency in the radio band, the local correlations do not hold spatially below certain scale 
lengths \citep{BASU2012}. This is due to cosmic ray electron propagation during their lifetime from their origin 
\citep{BASU2012}. High resolution radio continuum observations in nearby galaxies are needed to identify the right model 
for the propagation. The ratio of the radio to FIR emission is a measure of turbulent field amplification at present in the 
star forming regions \citep{TABATABAEI2013}. This ratio changes depending on the star formation rate and magnetic 
field strengths across the disk of a galaxy \citep{HEESEN2014}. High resolution sensitive observations of nearby galaxies 
could determine the level of turbulent field amplification in  different environment \citep{BECK2015B}. Tight correlation 
between non-thermal radio emission and molecular line emission (CO) has also been observed for galaxy M51 
\citep{SCHINNERER2013}. Coupling between molecular cloud density and non-thermal emission could arise due to 
(i) coupling between gas density and total magnetic fields, or (ii) increased synchrotron emission from secondary 
cosmic ray electron produced from interaction of cosmic rays with dense molecular material \citep{MURGIA2005}. 
However, the above observation was done with a spatial resolution of 60 pc, and observations with higher resolution in 
radio with SKA1 and molecular cloud with ALMA for for a larger sample of galaxies could differentiate between the 
above two scenarios \citep{BECK2015B}.

\subsubsection{Magnetic fields in galaxy halos}

Galaxies can show significant outflows to the intergalactic medium and can transport magnetic fields out of the 
galaxy disk \citep{HEALD2012}. Models of galactic outflows can reproduce the available multi frequency 
observations only if cosmic rays and magnetic fields activate the wind at their bases in galactic bulges 
\citep{EVERETT2008,Samui2010}. Earlier galactic wind models neglected cosmic rays, but, the energy density 
in cosmic rays is comparable to thermal gas, magnetic fields and turbulent motion energy density in the ISM 
\citep{BECK2015B}. Like our Milky Way, many galaxies with moderate star formation activity and supernovae 
can drive a galactic wind \citep{BREITSCHWERDT1991}. If the wind primarily consists of hadronic particles 
(e.g., protons), the $\gamma$-ray emission from it will show the typical pion bump in the GeV energy range 
and secondary electrons produce inverse Compton scattering. However, in case the wind is driven by leptons 
(e.g., electrons), inverse Compton emission from primary cosmic ray electrons will dominate in the GeV to TeV 
range. In radio bands, the synchrotron spectrum and polarization would have certain specific signatures from 
any of the above two processes \citep{BECK2015B}. Radio and gamma ray observations would provide important 
information on the origin and propagation of the cosmic rays and the magnetic field that affects their propagation. 
Secondary electrons are thought to be produced in a more isotropic environments that the primary electrons, and
is expected to show less polarized emission, which could be checked with sensitive SKA1 observations at GHz
frequencies \citep{BECK2015B}.  The Sunyaev-Zeldovich effect up scatters the cosmic microwave background 
photons after interactions with the particles in the wind which is detectable with sub-mm telescopes. The ratio of 
synchrotron to inverse scattered radiation can be used to measure the magnetic field energy in the wind 
\citep{BECK2015B} particularly for the edge-on galaxies. 

Cosmic ray electrons can propagate away from a galaxy by diffusion, streaming instability along magnetic field lines 
or by convective transport by a wind. As the electrons propagate away from their place of origin, their emission and 
spectrum changes with time. With high resolution observations of vertical profiles of radio emission and its spectrum, 
it is possible to determine the speed of cosmic ray electrons. This has been
done for NGC253 \citep{HEESEN2009} and more recently for a set of 35 galaxies
\citep{WIEGERT2015}. The latest observations with the JVLA indicate the regions
just outside the galaxies when averaged show presence of a halo of magnetic
fields and cosmic rays around them.
Clearly, SKA-I with an order of magnitude higher
sensitivity can bring out the
the size and shape of such halo around each of the individual galaxies. This
will allow to verify if a single model can explain the cosmic ray propagation or outflow
in different galaxies, or are there multiple processes (e.g., diffusion vs
streaming instability and outflow) and depending on other physical processes in
galaxies, one/some of them mostly influences the propagation of cosmic rays in
galaxies.

Polarization observations in nearby edge-on spirals show an X-shaped field in the halo \citep{KRAUSE2014}. 
Dynamo model of mean magnetic fields predicts poloidal fields in the halo. As in \citet{BECK2015B}, the theoretical 
estimate is much smaller than the measured values and this type of field structure can be created if the effects of 
galactic wind are included in model \citep{HANASZ2009}. RM grid observations with SKA1 can show whether the 
halo field is regular or composed of loops.

 \citet{SHNEIDER2014} presented a detailed calculation of the physics of depolarization of synchrotron radiation 
in the multi layer magneto-ionic medium in galaxies. Faraday tomography with high angular resolution, sensitivity and 
large bandwidth of SKA1-MID  would be useful along with RM grid to infer the morphology of the galactic magnetic 
field for galaxies to a distance of more than 10 Mpc with a spatial resolution of 100 pc \citep{HEALD2015}.

\subsubsection{Comparison of disk magnetic fields observations with theoretical models:}

In the local Universe, magnetic fields have been observed in more than a dozen nearby spirals \citep{vanEck+15}. 
These observations have revealed fields of several $\mu$G strengths ordered on kilo-parsec scales along with a 
random component with coherence scales of a few tens of parsecs \citep{Fletcher10, Beck16}. It is believed that 
dynamo model of amplification and maintenance of magnetic fields in galaxies could match the field strengths and 
its direction. However, strength of magnetic fields predicted from theory depends on several poorly measured galactic 
parameters. Compared to field strength, magnetic pitch angle, $p$ is a readily observable quantity whose 
measurements in different nearby spirals can be compared with predictions from the galactic mean field dynamo 
theory. The pitch angle is a measure of how tightly wound the large-scale field is and is expressed as, 
$\tan (p) = \overline{B_r}/\overline{B_{\phi}}$, 
where $\overline{B_r}$ and $\overline{B_{\phi}}$ are the radial and toroidal components of the mean magnetic field, 
respectively. If the field is simply frozen into the gas, it will get tightly wound by the galactic differential rotation leading 
to very small $p$. On the other hand, turbulent diffusion allows the field to partially slip through the gas, but then will 
lead to its decay, unless a mean-field dynamo operates to maintain the large-scale magnetic field. The observed values 
of $p$ are found to be larger than what is predicted from the galactic mean-field theory. For example, the pitch angles 
computed from the data set of \citet{vanEck+15} have a mean value of 25$^{\circ}$ with minimum and maximum 
values of $- 8^{\circ}$ and $- 48^{\circ}$, respectively. On the other hand, theoretical estimates from the standard 
$\alpha - \omega$ dynamo predicts $| p | < 15^{\circ}$ in the nonlinear regime. Attempts to resolve this mismatch has 
led several authors to explore different recipes \citep{CT15}, that range from making the disk thinner, the shear
smaller, assuming the kinetic $\alpha$-effect to be enhanced in the spiral arms \citep{MS91,CSS13}, using additional
helicity fluxes \citep{VC01, SSS07}, incorporating mean radial flows \citep{MSS00} to invoking spiral shocks \citep{vanEck+15}.
However, a limitation of the existing observational data is that different galaxies have been observed with different 
telescopes at different resolutions and frequencies. These complications directly hamper efficient comparison between 
theory and observations. The problem can be alleviated by new surveys with SKA at fixed resolutions and sensitivities. 
On the other hand, recent developments in galactic dynamo theory where the steady state of the mean magnetic field 
is controlled by magnetic helicity balance \citep{SB06}, predict that both the strength of the mean magnetic field and the 
pitch angle depend on the gas surface density $\Sigma_{\rm g}$ and the intensity of the outflow from the disk and 
therefore on the star formation rate (SFR) \citep{SSB06, vanEck+15}. 
Such nonlinear galactic dynamo models should also strive to incorporate realistic gas flows to improve agreement with 
observational data. 
The vast sensitivity of SKA will offer a unique possibility of tracing the total star formation rate in galaxies much better 
than those from existing telescopes \citep{Jarvis+15}. Thus, apart from improving theoretical models, a potentially robust 
measurement of the star formation rate in galaxies can lead to improved observational estimates of the magnetic pitch 
angle in the near future.

While the total magnetic field is observed to be stronger in the material spiral arms, quite surprisingly, in most of the 
observed spirals, the strongest ordered fields are detected in regions in between the material arms where the gas 
densities and turbulence are expected to be weaker \citep{Beck16}. A classic example of this are the pronounced 
{\it magnetic arms} in NGC6946. This constitutes a significant deviation from axial symmetry as one would expect the 
fields to be stronger in the material arms. It is rather difficult to explain such peculiar features by simply appealing to 
stronger turbulence or enhanced Faraday depolarization within the spiral arms. On the other hand, the presence of 
such non-axisymmetric features could be indicative of non-trivial interaction between the material spiral arms 
and the large-scale dynamo action. 

Over the years, several different explanations have been put forward to explain the high degree of 
coherence observed in the inter-arm regions. These range from invoking slow and fast MHD density 
waves in thin galactic disks \citep{LF98} (but using very simplified configurations for the magnetic field 
and galaxy rotation curves), more efficient dynamo action in the inter-arm regions due to a weaker 
$\alpha$-effect in the material arms \citep{M98}, drift of the magnetic fields w.r.t the gaseous arms 
caused by spiral perturbation \citep{OESU02} (kinematic model neglecting back reaction of the fields 
on the flow), stronger turbulence in the gaseous arms due to higher star formation rate leading to 
the mean-field saturating at a lower level \citep{S98,Moss+13}, introducing a time delay of the magnetic 
response in the mean-field equations \citep{CSS13}, to weakening of the mean-field dynamo in the 
material arms by star-formation driven outflows \citep{CSS15}. 

Among these, the work of \citet{CSS13} explores these non-axisymmetric features by including a 
finite time delay in response of the mean electromotive force to changes in the mean-magnetic field and 
small-scale turbulence in a non-linear dynamo model based on magnetic helicity balance. This leads to 
a phase-shift between the material and magnetic spiral arms which reproduces the kind of features observed 
in NGC6946.
However, the non-axisymmetric components of the mean-magnetic field driven by the spiral 
pattern are mainly localised around the co-rotation radius. 

All these different approaches indicate that our understanding about the occurrence of magnetic spiral arms 
is nascent, and future research directions would require a synergistic approach between theory, observations 
and numerical simulations. Recent work on galactic spiral structure \citep{Dobbs+10, Quillen+11} has revealed 
that the spiral patterns of many galaxies appear to wind up as opposed to being rigidly rotating (as is usually 
assumed in galactic dynamo models). Galactic dynamo models incorporating such winding up spirals \citep{CSQ14,CSS15} 
have led to magnetic arms spread over a wider range of radii, and so matching observations better. This suggests
an interesting synergy between dynamo theory and spiral structure theory which needs to be explored further. 
On the observational front, detailed high resolution data of diffuse polarized emission and Faraday rotation measure 
(as can be obtained from SKA-mid) can help us in understanding such magnetic arms better. Moreover, as pointed by 
\citet{Beck+15}, the interaction between density perturbations and magnetic fields can be probed in great detail with 
data of neutral and ionised gas obtained from SKA1. Besides, it would also be useful to obtain constraints on the 
rotation curve, velocity dispersion and spiral structure from existing data.

{\large \bf Alternative dynamo models -} There exist dynamo models based on instabilities like the Magneto-rotational
instability (MRI) to explain the generation and maintenance magnetic fields in the extra stellar disk of galaxies. In an 
earlier work \citep{BASU2013}, the magnetic field energy density in the outer parts of the galaxies were found to be 
higher than the turbulent total gas energy density. Mean field dynamo model may not generate such fields. 
\citet{Sellwood_Balbus99} have shown that the existence of the high H~{\sc i} velocity dispersion in the extra
stellar disk of the spiral galaxies can be produced by extraction of the energy from galactic differential rotation through 
magneto-hydro dynamic turbulence. If MRI causes turbulence at large galactic radii, an associated $\alpha$ effect
may develop, and regular fields will also be seen out to such large radii \citep{PRASAD2016}. A much larger sample 
size than discussed above will help to confirm such a conclusion. RM grid observations with SKA1-MID would be the 
right choice to pursue such a project.

\subsection{Magnetic fields in interacting galaxies}

Interaction of galaxies form bridges between them as matter flows from one system to another. This matter, if in the 
form of a plasma, would carry magnetic fields due to (at least partial) flux freezing. Thus magnetic bridges could be 
additional probes to study galactic interactions \citep{DRZAZGA2011, DRZAZGA2012,Beck16}. 
In fact one could envisage a situation where the material bridge although present is not visible but magnetic field 
structures could be detected by Faraday RMs in the intergalactic space between two galaxies. This could form a 
new probe to detect and~study galaxy-galaxy interactions.  To draw generic features of interactions, one would need 
several systems where the interaction is detectable through enhanced FDs of background sources. Further, one needs 
to study the detailed structure of the magnetic fields in the interacting region. In fact the existence of possible 
small-scale magnetic field structures in interacting galaxies could produce both rotation of the polarization angle and 
depolarization. This will cause an enhancement of the FD variance, $\sigma_{FD}$, in the line of sight having intervening 
interacting galaxies.
SKA-I would prove to be~indispensable in this 
context. Polarised background sources seen through the interacting region will be able to probe such regions 
\citep{AKAHORI2014A, AKAHORI2014B}. Assuming two galaxies in the nearby universe separated by $\sim 20'$ and 
the bridge between them to have a width of 5$'$, with an integration time of 1 hour with SKA-I mid, $\sim 30$ 
background sources with measurable FDs will be observed through the region. 
Hence, the need of finding radio sources behind these systems is almost always satisfied. 
Comparing their $\sigma_{FD}$ with the
$\sigma_{FD}$ of 
sources seen through the surrounding regions shall detect enhanced magnetic field strength due to matter drawn from
these galaxies.
Further, the 
resolution of the SKA will be high enough to~analyse~the structure of the bridges.

\subsection{Turbulence in the interstellar medium and galactic dynamo}

In the standard $\alpha - \omega$ dynamo picture, the toroidal component of the magnetic field is generated from 
the radial component through differential rotation, while the radial component is regenerated from the toroidal component 
through the $\alpha$-effect, involving helical turbulent motions in the disk. In simple terms, the growth of the large-scale 
fields is controlled by the dynamo control parameters, $R_{\alpha} = {\alpha h}/{\eta_t}$ and
$R_{\omega} = Gh^2/\eta_t$ characterising the intensity of induction effects due to helical turbulence and 
differential rotation respectively. Here $h$ is the disk scale-height, $G$ is the large-scale velocity shear rate, 
$\eta_t = ul/3$ is the turbulent magnetic diffusivity and $\alpha \simeq {l^2 \Omega}/h$ with $l$ being the driving 
scale of interstellar turbulence and $\Omega$ is the angular velocity of rotation. Star formation in spiral galaxies, 
resulting in supernova explosions are the main drivers of turbulence in the ISM. Large-scale magnetic fields in galaxies 
are produced by motions in  the diffuse, warm gas which is partially ionized. In most studies of galactic dynamos, the 
driving scale is chosen to be $\sim$100 pc and the typical turbulent velocity, $u\sim 10$ $\tmop{km} {\rm s}^{- 1}$ 
which is equal to the sound speed of the gas in the warm medium. However, if supernovae go off randomly in space, 
it is difficult to justify a particular value for the driving scale of turbulence. In fact, numerical simulations of supernovae 
driven turbulence in a stratified medium by \citet{JM06} show that there is no single effective driving scale. Instead, the 
kinetic energy is distributed over a wide range of wave numbers. Note here that the turbulent driving scale and the turbulent 
velocity also determine the amplitude of the turbulent diffusivity. Moreover, both the driving scale and the turbulent velocity can 
vary depending on the gas surface density and the supernovae rate in the galaxy. In fact, observations by \citet{Genzel+11} 
and \citet{Swinbank+11} find velocity dispersions of $\sim 50 - 100 \tmop{km} {\rm s}^{- 1}$ in high surface density disks, 
which could possibly arise from gravitational instabilities \citep{SSO16}. Thus, it is crucial to make correct estimates of 
these quantities which in turn will provide improved estimates of the dynamo parameters. However, the nature of 
turbulent flows in the magnetised ISM is largely unknown as the key properties of turbulence are poorly constrained 
by observations. Recent work by \citet{Gaensler+11} and \citet{BLG12} has shown that the skewness and kurtosis
of polarization gradients of synchrotron emission can constrain the sonic Mach number in the warm ionized medium. 
Correlation of the Mach number with the temperature can provide insights into the nature of the turbulent cascade in
the ISM. Such techniques coupled with statistical analysis of synchrotron intensity fluctuations can provide robust 
estimates of the turbulent parameters in the ISM. These measurements could compliment the direct detection of 
turbulent velocities through observations of the redshifted 21 cm line. One will require a signal-to-noise ratio of 30, an 
angular resolution of about 20$''$ and a sensitivity of 1 $\mu \tmop{Jy} \tmop{beam}^{- 1}$ \citep{Herron+16}, which 
can possibly be achieved with the SKA.

A concomitant issue is the spectrum of the small-scale magnetic fields which forms the turbulent component
of the total magnetic field. Such fields could result from either the operation of a small-scale dynamo in the galactic 
ISM or from the tangling of the regular magnetic field. Both of these components give rise to synchrotron 
emission in the sky plane.
Thus, how can one distinguish between these two types of fields observationally, in terms of their power spectra 
and correlation lengths?
If such fields arise from small-scale dynamos, they are likely to be less volume filling compared to their tangled 
counter parts.
However, given the limitations in spatial resolution of present day radio observations, it is difficult to 
clearly distinguish the typical correlation length scales, and the power spectra of these components. 
In fact, power spectra of a tangled magnetic field is expected to show fluctuations over a wide range of 
spatial scales as the magnetic field is tangled on multiple different scales. 
If the turbulent component of the field is generated only by small-scale dynamo action, one would expect them to be 
correlated on at most the scale of turbulence in the galaxy. Their presence can be identified from large fluctuations 
in the synchrotron emission while the turbulent component resulting from tangling of the regular field is expected 
to give rise to moderate levels of fluctuations in synchrotron emission. In a recent work, \citet{Houde+13} derived 
the degree of anisotropy in magnetic field fluctuations from the scatter in the observed polarization angles at 
high radio frequencies (i.e., for small Faraday rotation angles). But as pointed by \citet{Beck+15}, this technique 
is now restricted to only a few bright patches of polarization seen at the highest possible resolution in M51 with 
the correlation lengths parallel and perpendicular to the local ordered field being about $100$ and $50\,{\rm pc}$ 
respectively. 

Given the high spatial resolution offered by the SKA, it would become possible to resolve these issues through 
polarization observations at spatial resolutions of $1-100\,{\rm pc}$. Moreover, improved Faraday depolarization 
measurements may also shed valuable information on the correlation scale.

\subsection{Probing magnetic fields at high redshifts}

MgII absorption systems probed by \citet{Bernet+08, Bernet+10,Farnes+14} also reveal the existence of magnetic fields in 
galaxies at redshifts $z \sim 1$. These fields are of comparable strengths to those that are observed in galaxies 
of today. This raises the interesting question of how such strong magnetic fields are generated at early epochs 
and how such fields evolve in redshift along with the evolution of the galaxy itself. 
Specific predictions about magnetic fields in young galaxies were presented in \citet{Arshakian+09} and 
\citet{SB13}. 
In the context of hierarchical galaxy formation, magnetic fields have been considered by \citet{RODRIGUES2015}. 
Detailed study of field evolution at high redshift requires one to explore how major mergers, star-formation rate (SFR) 
etc. influence the gradual evolution of magnetic fields with time. While the fields in these galaxies may not have had 
the time to get ordered on larger scales (i.e., on scales bigger than the scale of turbulence), they can nevertheless be 
detected via their resulting synchrotron emission. However, while energy losses of cosmic ray electrons due to inverse 
Compton scattering off CMB photons are negligible in the local Universe, the rapid increase of the CMB energy density 
by a factor $\sim (1 + z)^4$ indicates that at high redshifts, such losses will become dominant compared to losses due 
to synchrotron radiation  \citep{Murphy09}. This in turn will make it difficult to observe galaxies in radio beyond a 
certain critical redshift. Deep imaging capabilities of SKA-I mid at GHz frequencies can be used to probe the
polarization properties of galaxies at high redshifts. In particular, SKA-I is expected to detect $\sim$5000 galaxies 
per square degree above 10 $\sigma$, which implies that it can probe magnetic fields in $\sim$50,000 galaxies 
out to redshift $z > 4$ for a 10 square degree survey \citep{Taylor+15}. This will allow to test different models which 
could generate magnetic fields in short timescale.
 
These include (i) fluctuation dynamos, (ii) quasi-linear global dynamo models with specific radial forms for diffusivity 
and alpha effect, (iii) models which treat growth of fluctuations and mean in a unified manner \citep{BHAT2016}.
Fluctuation dynamos, appear to be a suitable candidate capable of rapid amplification of magnetic fields on the 
eddy-turnover-time scale, much shorter than the lifetime of a galaxy \citep{BS05}. They also can lead to sufficiently 
coherent fields to explain the observations \citep{BS13}. Such dynamos can also generate fields much before the
formation of the disk by tapping into the energy in the turbulent motions of the halo gas as the galaxy forms 
\citep{Sur+10, Sur+12}.

Mean field models with helicity fluxes can avoid catastrophic quenching of the dynamo, and the simplest example 
of such a flux arises from advection of the field through galactic outflows \citep{SSB06}. In the case of quasi-linear 
dynamo model (Prasad \& Mangalam, 2016), a global axisymmetric turbulent dynamo operates in a galaxy with a 
corona. It treats the supernovae (SNe) and magneto-rotational instability (MRI) driven turbulence parameters under 
a common formalism where the diffusivity and alpha effect have radial variation as the shear. The nonlinear quenching 
of the dynamo is alleviated by inclusion of small-scale advective and diffusive magnetic helicity fluxes (eg. \citet{SSS07}, \citet{CHAMANDY2014}, which allow the gauge invariant magnetic helicity to be transferred outside the disc and
consequently build up a corona (halo) during the course of dynamo action. 
The quadrupolar large-scale magnetic field in the disc is found to reach equipartition strength within a timescale of 
1 Gyr which is much smaller than those predicted by other models. The large-scale magnetic field in the corona
obtained is much weaker in strength compared to the field inside the disc and has only a weak impact on the dynamo 
operation.

The structure and strength of the 
coronal magnetic field is one important discriminant. Also, observations of the galactic magnetic field strength at different 
redshifts is key to understanding the timescale of saturation of the dynamo. In addition the magnetic pitch angle obtained 
by this model can be compared with observed values.

\section{Magnetic fields in galaxy clusters}

Galaxy clusters are found to be magnetised and these fields are important in understanding the physical 
processes in the intra cluster medium is also recognised. Several large scale (few Mpc) features from the clusters 
have already been detected \citep{BAGCHI2002} with surface brightness of the order of few tens of $\mu$Jy/arc-sec$^2$ 
at 1.4 GHz. Central region of clusters often host diffuse radio halos of non thermal synchrotron
emission \citep{GOVONI2004} with intensity strongly correlated with the X-ray luminosity, and hence the mass of 
the cluster. Recent simulations of \citet{GOVONI2013} suggest that the halos have intrinsically polarized
emission which can be traced when observed at high resolutions (100 pc or lower). 
SKA1-Mid at 1.4 GHz with a resolution of about 0.2 arc sec would have the right polarisation sensitivity to observe 
these in many clusters \citep{GIOVANNINI+15}. 

While synchrotron radiation traces the component of the magnetic field perpendicular to the line of sight of observation, 
RM synthesis traces the line of sight component. Using models of the magnetic fields it has been shown that this 
technique can be used against background radio galaxies for clusters with mass $> 10^{13} M_{\odot}$
\citep{BONAFEDE2015}. Moreover, this method would be effective in probing the compressed magnetic fields at the 
shock fronts from the merging clusters. 
\citet{JOHNSTON-HOLLITT2015} outlined a technique to investigate the evolution of the intra cluster magnetic fields 
over cosmic time using their imprint on the tailed radio galaxies. Probing details in the morphology of the
tails require high sensitivity and resolution.

Magnetic fields in clusters are possibly generated by the Fluctuation dynamo 
which amplifies magnetic fields on the fast eddy-turnover time scales and on coherence lengths smaller than the 
outer scale of turbulence. A crucial issue in fluctuation dynamos is the degree of coherence of the field in the saturated 
state. A recent study of directly measuring the RM from simulations of the fluctuation dynamos does seem to indicate 
the generated fields are coherent enough \citep{BS13}. The SKA will enable the detection of many more polarized 
sources through an individual cluster and thus map the random field in it. One would also detect many more radio 
halos. In particular the enhanced sensitivity and improved angular resolution of SKA will allow one to detect polarization, 
which has been seen in very few radio halos at present \citep{Govoni15}. The use of a wide bandwidth could also allow 
one to do RM synthesis and infer the 3-d structure of the magnetic field. The detailed mapping of the continuum emission 
including its fluctuations will also probe the coherence properties of the magnetic field and the cosmic ray electrons.

Away from the cluster cores, we have very little information at present about the magnetic 
field strength and their distribution in the cosmic web, particularly in the intergalactic filaments. Magnetic 
fields in filaments can be amplified by vorticity and turbulence resulting from structure formation shocks 
\citep{Ryu+08, Ryu+12, ISNM11}. On the observational front, apart 
from faint radio emission observed in the outskirts of clusters \citep{Kim+89}, there is no confirmed 
detection of synchrotron emission and RM from filaments. Based on the observed limit of RMs of background 
quasars, \citet{RKB98} and \citet{Xu+06} inferred an upper limit of $\sim 0.1\,\mkG$ for the magnetic field 
strength in filaments. With the growth of computing power and availability of sophisticated numerical algorithms, 
it has now been possible to perform cosmological MHD simulations of large-scale structure formation including 
the formation of galaxy clusters. A major drawback of these simulations is their inadequate resolution to resolve 
the turbulent eddy scale in filaments. This makes it difficult to reach firm conclusions 
regarding the degree of turbulence required to amplify the magnetic fields and whether the fields in the filaments 
are in equipartition or are still evolving. Moreover, the field amplification is found to depend on the numerical 
resolution as well as on the distribution of solenoidal or compressive modes of the underlying turbulence 
\citep{Xu+12, Vazza+14}. Nevertheless, they offer a useful first hand estimate of the magnetic field strengths 
in the intergalactic filaments. 

An interesting approach adopted by \citet{Ryu+08} combined the estimated levels of resulting turbulence in a 
large-scale structure simulation with the magnetic field growth due to fluctuation dynamos obtained from a separate 
three-dimensional incompressible simulation of driven turbulence. This gives an estimate of the magnetic field of tens 
of $\nG$ in these filaments and their RM contribution to be $\sim 1\,{\rm rad\,m}^{-2}$. Again, the crucial question is 
the degree of coherence of the field. Apart from the fact that measuring RMs in these filaments are beyond the reach 
of current observational facilities, one is also confronted by difficulties arising out of separating RM contribution due
to other sources of Faraday rotation along the line of sight (LOS). These include FDs arising from : other background 
extragalactic radio sources, other intervening galaxies that may lie along the LOS, the magnetic field in our Milky 
Way and RMs associated with the earth's ionosphere. The contribution to the RM from these sources are not 
negligible compared to the FDs expected from the magnetic field in the filaments. Thus, one has to devise 
a clever way of separating the FD contribution from these other sources from the RMs arising solely from 
filaments and large-scale structure. In this context, new statistical techniques such as in \citet{AKAHORI2014A}, 
Faraday tomography measurements \citep{AKAHORI2014B} and measuring both the dispersion measure (DM) and 
FD of extragalactic linearly polarized fast radio bursts \citep{ARG16} hold promise for unravelling the strength and 
FDs of magnetic fields in filaments with the SKA. 
Again as pointed out earlier, the complicated internal polarization structure of the background AGN's
need to be understood \citep{AGF16, Kim+16}, before 
before being to able to measure the RM contribution from 
the IGM. 
On the theoretical front, the key challenge lies in understanding 
the saturation process of the fluctuation dynamo, for which higher resolution simulations with high fluid and magnetic 
Reynolds number at magnetic Prandtl number ${\rm Pm} \gg 1$ are needed.

\section{Magnetic fields in AGN parsec scale jets: implications for black hole physics}

Blazars are core dominated Active Galactic Nuclei (AGNs) and are characterised with luminous core, rapid
variability over entire electro magnetic (EM) spectra, high radio to optical polarization, superluminal motion, non 
thermal emission and a doppler boosted relativistic jet pointing $\leq$ 10$^{\circ}$ with the line of sight (LOS).  
Core-jet morphology is common characteristic of most of the Active Galactic Nuclei (AGNs) in very long baseline 
interferometry (VLBI) images, where core is the optically thick base of the jet \citep{BlandfordKonigl1979} with its 
absolute position being the surface in the continuous flow where optical depth becomes $\sim$ 1 (also known as
'photosphere').  

According to \cite{Konigl1981}, absolute position of the VLBI core moves increasingly outwards along the relativistic 
jet with higher wavelength which can be attributed to the synchrotron self absorption process (SSA), calling this effect 
as frequency dependent core shift.   In the conical jet model, the 
VLBI core is a compact, stationary, bright and flat spectrum feature lying at one end of the jet of a typical blazar on 
sub-milliarcsecond (pc) scale from where super luminal components emerge and separate which appear as propagating 
disturbances such as shock waves \citep{BlandfordKonigl1979}. In case of blazars the spectra of the cores at centimetre 
wavelength indicates core is partially optically thick to the synchrotron self absorption since their slope is $\leq$ 1, which 
can be attributed to magnetic field and the relativistic electron density gradients. 

High precision core shift measurements that can be done with VLBI involving
SKA with other large area telescopes can prove to be very fruitful as it can potentially provide information on
the physical parameters of the VLBI jet, like core magnetic field, distance
from the core to the 
base of the jet \citep{Lobanov1998, Hirotani2005}, spectral index, construction of rotation measure maps, and 
nature of the absorbing material with SSA being dominant in the jet plasma while free free absorption driving the thermal
plasma surrounding the jet in the sources viewed at large angle to the line of sight (LOS). There are several ways to 
measure apparent core-shifts. One of them is the phase referencing experiment where the telescope is switched
between the target source and the phase calibrator (some nearby reference source) with shorter switching time as 
compared to the coherence time \citep{Guirado1995, Marcaide1984, Lara1994, Bietenholz2004}. A new technique is 
used for extracting core shifts along with other physical parameters of jet using frequency dependent time lags for 
single dish observations (\citet{Mohan2015}; Agarwal et al. in preparation). The core-shift effect is used effectively to 
probe the region close to the jet launching region by a synchrotron opacity model. Applying the model to multi-band 
observations (4.8 GHz - 36.8 GHz) including high resolution VLBI images and light curves, they obtain the component 
kinematics and in the context of the core-shift effect, to constrain the core radius, magnetic field strength (core and at 
pc-scales), core offset position and jet luminosity, assuming a steady conical jet affected by core brightening (flares) 
and associated component ejection.

Magnetic fields are expected to play a prominent role in structuring the accretion flow, the disk-jet connection, and 
the collimated outflow at the sub-pc to pc-scales and in possible helical signatures at the kpc-scales. Further, using 
the derived field strength and jet luminosity one can explore electrodynamical and hybrid jet models and place constraints 
on the spin and mass of the black hole. \citet{Z14} considered the samples of blazars and radio galaxies of and obtained 
core shift and magnetic field strengths. Theoretical results for equipartition and the jet opening angles can be tested. 
\citet{Z14} considered the model of jet formation from black hole (BH) spin-energy extraction \citep{BZ77}. \citet{Z15} 
considered models with the accretion being magnetically arrested (MAD; \citet{Narayan2003}). 
Such flows have dragged so much magnetic flux to the BH that the flux, $\Phi_{BH}$ becomes dynamically 
important and obstructs the accretion, $\Phi_{BH} \simeq 50 (\dot{M} c)^{1/2} = 10^{-4} M_8^{3/2} \dot{m}^{1/2}$ pc$^2$ G
\citep{Tchek2011, McKinney2012} where the accretion rate is given in units of the Eddington rate.
Further, using the derived field strength and jet luminosity one can explore other electrodynamical 
and hybrid jet models and place constraints on the spin and mass of the black hole.  
SKA as part of a VLBI network with long baselines can provide key and unique contributions to the problem:

\begin{enumerate}
\item by verifying the theoretical SSA model using high resolution maps on sub pc scales and the determination of the
optical depth as a function of radius is key to measuring the magnetic fields.

\item by  spanning the range 50 MHz to 14 GHz and  monitoring these AGNs for several years,
 we will be
able to broaden the scope for such studies and offer a sizable sample for
statistical and comparative studies of jet kinematics and energetics at the pc-scale which can serve as crucial
inputs to fully relativistic Radiation MHD numerical simulations.
\item by going deeper in redshift, exploiting the higher sensitivity has important cosmological implications for evolution of magnetic fields in these
AGNs
\end{enumerate} 

\section{Radio observations of Solar magnetic fields}

Solar coronal magnetic field can be effectively estimated using low frequency radio observations ($< 150$ MHz or so) 
in the cross-correlation mode between signals received by two mutually orthogonal dipole antennas \citep{Sastry2009}. 
It is only the circularly polarized radio emission that is observed since linearly polarized radio emission averages to zero 
over typical observing bandwidths at low frequencies.

Presently such observations are carried out in India at the Gauribidanur observatory with an one-dimensional array of 
antennas on an east-west baseline. Several interesting results on the coronal magnetic field (particularly when there is 
a coronal mass ejection) in the heliocentric distance range $\sim 1.2 - 2.0$ solar radii have been obtained \citep{Ramesh2010}.
Large antenna arrays like SKA are expected to be more useful in this regard since:

\begin{itemize}
\item The higher angular resolution will help to locate the radio wave emitting source regions unambiguously. 
This will facilitate comparison with observations at other wavelength bands in the spectrum, both laterally and
radially. Note that simultaneous observations at different frequencies help to derive the radial variation of the field 
strength.

\item The larger collecting area (and hence better sensitivity) will help to observe faint radio emission. Such 
observations will be useful to estimate the coronal magnetic filed associated with weak energy releases in the 
solar atmosphere. Radio emission from background corona can also be probed to infer the general magnetic field 
there.
\end{itemize}

Given this we consider two specific proposals below.

\subsection{Topological properties of solar coronal fields derived from nonlinear force-free (NLFF) magnetic field solutions}
\citet{PRASAD2014} conducted a systematic study of force-free field equation for simple axisymmetric configurations in 
spherical geometry. The condition of separability of solutions in radial and angular variables leads to two classes of 
solutions: linear and non-linear force-free fields. They have studied these linear solutions and extended the non-linear 
solutions to the irreducible rational form $n = p / q$, which is allowed for all cases of odd $p$ and to cases of $q > p$ for 
even $p$. They have further calculated their energies and relative helicities for magnetic field configurations in finite and 
infinite shell geometries. They demonstrate a method to be used to fit observed solar magnetograms as well as to provide 
good exact input fields for testing other numerical codes used in reconstruction on the non-linear force-free fields. They 
have further calculated their energies and relative helicities for these magnetic field configurations in finite and infinite
shell geometries. This method can provide useful reconstruction of the non-linear force-free fields as well as reasonably 
good input fields for other numerical techniques. Observations of the solar magnetic fields by SKA can be used inputs to 
test the solutions and the predicted topologies.  A  set of solutions including pressure obtained by solving the Grad Shafranov 
can also be tested: self-similar solutions with twist \citep{Osh1982}, non-self similar without twist  \citep{Low1975a, Low1975b} 
and more generally with twist (Sen \& Mangalam, in preparation).

\subsection{Energy distribution of braided fields in Sun derived from NLFFF solutions}

Observations of solar coronal loops reveal highly regular structures. It is expected that random rotations and random 
walk of the foot points at the photosphere twist and braid the field lines, then how can the random processes existing 
in corona lead to such well organised structures? Several observations have reported evidence of such braiding of 
magnetic field lines. It is possible that the magnetic re-connection within the loop and with other loops disentangle the 
field through nanoflares or microflares. The coronal field then reorganises itself through a Taylor like relaxation process 
to attain a force-free field configuration. It is useful and important to study this reorganisation process of a highly braided 
field to a force-free field geometry. The idea of crossing numbers and similar topological constructions are used to calculate 
the free energy content in braiding in the field lines which allows estimates of the energy released in flares from these 
structures. In a paper in preparation, \cite{PRASAD2016} test the statistical model of self-organised criticality using their 
nonlinear force-free field (NLFFF) solutions to find the distribution function of crossing numbers and the power law in the 
energy distribution of flares. Their results are in good agreement with those predicted in the model by 
\citet{PRASAD2013, PRASAD2014}. Further observations by SKA on the flare distribution energies and more detailed 
estimates of the geometry of the braided structure will be directly relevant predictions of these models.

\section*{Conclusions}
Having broadly discussed the various topics of interest to the Indian community, we summarise here the specific 
objectives of our group which includes but is not limited to the following science cases :

\begin{enumerate}

\item RM grid observations of sources through some of the discrete objects in
   the Galaxy including the GC region, molecular clouds and SNRs (Sect. 3.2.2)
   to understand and model the magnetic field distribution in these objects and
   compare the effect of large scale magnetic fields in these objects.

\item 
   As discussed in Sect. 3.3.1, observing a set of spirals in high resolution to model the
   propagation of cosmic rays along the plane and to study its correlation with emission in other bands
   to determine the coupling of magnetic fields with other constituents of the ISM.

\item 
   Observe a set of nearby spirals (Sect. 3.3.2) with large bandwidths in
   polarization to determine and model the halo magnetic fields. Before the SKA survey band 2 is
   operational, we also plan to use the 0.6$-$0.9 and 1.0$-$1.4 GHz bands of
   GMRT to carry out RM synthesis observations of some of the nearby galaxies.

\item
   Comparing magnetic field
   observations of disk galaxies with theoretical predictions (Sect. 3.3.3) - improving 
agreement with observations in the values of the magnetic pitch angles and also improving 
our understanding of non-trivial interactions between material spiral arms and large-scale 
dynamo action. 

\item Probe interacting galaxies and the enhanced magnetic field strength (due to the matter drawn 
in from these galaxies), by carrying out the comparison of the variance of the Faraday depth, $\sigma_{FD}$, 
of radio sources sample with and without these intervening interacting galaxies along their line of 
sight (e.g Sect 3.4).

\item Nature of turbulent flows in the magnetised ISM (Sect 3.5) is largely unknown as crucial properties 
of turbulence are poorly constrained at present. Statistical analysis of synchrotron intensity 
fluctuations as already reported by a number of authors, along with direct detection of turbulent 
velocities through observations of the redshifted 21-cm line can provide robust estimates of the 
turbulent parameters in the ISM. These estimates can then be incorporated into nonlinear 
galactic dynamo models to provide more insights into the evolution of large-scale magnetic 
fields in galaxies. 

\item Probing the evolution of cosmic magnetic fields from high redshifts (Sect. 3.6) to the present
through both new observations and improved theoretical predictions for SKA and its precursors.

\item Probing the degree of coherence of the magnetic field in galaxy clusters. In particular, 
the use of wide bandwidth will allow one to do RM synthesis and infer the three-dimensional
structure of the magnetic field. Also, the detailed mapping of the continuum emission will be 
helpful to probe the degree of coherence of the magnetic field. 

\item
By studying the core-shift as a function of frequency over larger frequency and dynamic range in radio map,s
SKA can probe magnetic fields, kinematics and spectral distribution which are key inputs to the synchrotron 
self-absorption models, radiation MHD simulations and black hole spin and mass estimates based on GRMHD 
calculations (section 5). Monitoring several sources for several years provides cosmological evolution of sub 
pc properties such as magnetic field  evolution and kinematics which are important clues to jet physics in these 
sources.
 
\item
Magnetic configurations on the Sun extending from the photosphere to the corona can be studied and 
force-free and more general models can be tested (section 6.1). Furthermore observations by SKA on the 
flare energy distribution and the geometry of the braided structure will be directly 
relevant to predictions of the model based on braiding topologies and magnetic reconnections  leading to 
nano flares (section 6.2).

\end{enumerate} 

\section*{Acknowledgement}

We acknowledge valuable inputs by Prasun Dutta during the course of revision of the work presented here. 
We also thank Rainer Beck for his insightful comments on the prospects of measuring the RM contribution from 
the IGM with future radio telescopes. 

\bibliographystyle{apj}

\bibliography{magnetic.field.ska.india.astroph}

\end{document}